\documentclass[
 aps, pra,
 amsmath,amssymb,
 11pt,
 final,
tightenlines,
 twoside,
 twocolumn,
 nofloats,
nofootinbib,
 superscriptaddress,
showkeys,
showkeywords,
 ]
{revtex4-2}

\usepackage[T1]{fontenc}
\usepackage[utf8x]{inputenc}
\usepackage[english]{babel}
\usepackage{graphicx}
\usepackage{dcolumn}
\usepackage{bm}

\input{maik.rty}

\setcitestyle{authoryear,round}
\def\squareforqed{\hbox{\rlap{$\sqcap$}$\sqcup$}}

\def\sq{\ifmmode\squareforqed\else{\unskip\nobreak\hfil
\penalty50\hskip1em\null\nobreak\hfil\squareforqed
\parfillskip=0pt\finalhyphendemerits=0\endgraf}\fi}

\def\degr{\hbox{$^\circ$}}

\def\utw{\smash{\rlap{\lower5pt\hbox{$\sim$}}}}

\def\udtw{\smash{\rlap{\lower6pt\hbox{$\approx$}}}}

\def\diameter{{\ifmmode\mathchoice
{\ooalign{\hfil\hbox{$\displaystyle/$}\hfil\crcr
{\hbox{$\displaystyle\mathchar"20D$}}}}
{\ooalign{\hfil\hbox{$\textstyle/$}\hfil\crcr
{\hbox{$\textstyle\mathchar"20D$}}}}
{\ooalign{\hfil\hbox{$\scriptstyle/$}\hfil\crcr
{\hbox{$\scriptstyle\mathchar"20D$}}}}
{\ooalign{\hfil\hbox{$\scriptscriptstyle/$}\hfil\crcr
{\hbox{$\scriptscriptstyle\mathchar"20D$}}}}
\else{\ooalign{\hfil/\hfil\crcr\mathhexbox20D}}%
\fi}}

\begin{document}

\selectlanguage{english}



\title{Kinematics of the System of Young Open Clusters based on Gaia\,DR3 Data}

\author{\firstname{A. K.}~\surname{Dambis}}
\affiliation{Sternberg Astronomical Institute, Lomonosov Moscow State University,
	13, Universitetskii prospect, Moscow, 119992, Russia}
\author{\firstname{A. S.}~\surname{Rastorguev}}
 \email{alex.rastorguev@gmail.com}
\affiliation{Sternberg Astronomical Institute, Lomonosov Moscow State University,
	13, Universitetskii prospect, Moscow, 119992, Russia}
 \affiliation{Lomonosov Moscow State University, Faculty of Physics, 1, bld.2, Leninskie Gory, Moscow, 119992, Russia}

\begin{abstract}
We use a kinematical model including circular rotation of the Milky-Way disk incorporating effects produced by a spiral density wave
with constant radial and vertical components of the velocity dispersion tensor and adopted solar Galactocentric distance
$R_0$~=~8.277~kpc (inferred by GRAVITY collaboration in 2022) to derive the velocity field parameters for a sample of
2384 open clusters from the  \citet{HR} catalog having ages no greater than 100~Myr almost all of which reside at
Galactoaxial distances between  5 and 14~kpc. We estimate the velocity-field parameters using maximum-likelyhood method
and the affine-invariant variant of the Markov chain Monte-Carlo method proposed by  \citet{GW}
with both methods yieilding almost identical parameter values.  The inferred rotation curve, on the whole,
is consistent with the results based on our kinematical analysis of a sample of Galactic masers. We find the
linear rotation velocity at the solar distance to be  $V$~=~242.3~$\pm$~1.1~km/s; it reaches its maximum  $V\sim$~244~km/s at
Galactoaxial distance of about $R_g\sim$~7~kpc followed by a slow and smooth decline with $V\sim$~227~km/s at $R_g$=~14~kpc.
The inferred radial and vertical components of the velocity dispersion tensor
are ($\sigma U_0$, $\sigma W_0$)~$\sim$~(10.63~$\pm$~0.15, 4.56~$\pm$~0.15)~km/s. We find the pitch angle and the phase of the
Sun for a four-armed spiral pattern to be $i\sim$~-11.7~$\pm$~0.3$\degr$ and $\chi_0\sim$~138~$\pm$~5$\degr$, respectively, and
the amplitudes of radial and tangential perturbations,  $f_R$~=-3.0~$\pm$~0.4 and $f_{\theta}$~=-3.1~$\pm$~0.4~km/s, respectively.
\end{abstract}

\maketitle

\section{INTRODUCTION}

The publication of extensive high-precision astrometric and radial-velocity data in the third data release of the Gaia space
astrometry mission (Gaia DR3) \citep{GAIADR3} has made open clusters extremely efficient and precise mass distance indicators
and kinematic tracers. This is primarily due to the comprehensive study by \citet{HR}, who determined the parameters of over
7000 clusters in our Galaxy based on Gaia DR3 data (trigonometric parallaxes, proper motions, two-color photometry, and
radial velocities) for more than 1.2 million stars.The \citet{HR} catalog is the first such extensive sample of clusters with
homogeneous estimates of distances, ages, proper motions, and radial velocities — based solely on high-precision measurements of
trigonometric parallaxes, proper motions, radial velocities, and two-color photometry of individual stars adopted from
a single source. Age estimates for more than 2000 clusters in the \citet{HR} catalog do not exceed 100 Myr, and these young
entities represent an ideal tool for a detailed study of the kinematics of the young Galactic disc, including the Galactic
rotation curve as well as non-circular motions induced by spiral density waves. Previously, such comprehensive studies of young
disc kinematics had to be carried out using data on
Galactic masers \citep{Reid13, Reid18, Xin_Zheng, BobylevBajkova13, BobylevBajkova14a, BobylevBajkova14b,
Bobylev_etal25a, Bobylev_etal25b, Reid14, Reid19, Rastorguev17, ImmerRygl, Nakanishi, Nikiforov} whose chief drawback is
the small size of the available samples with the relevant measurement data, numbering at most about two hundred masers.
The sample size of young clusters from \citet{HR} is an order of magnitude larger with comparable precision of individual
distance and proper motion estimates and a comparable spatial coverage (with heliocentric distances reaching  10 kpc and more).

Just recently \citet{BobylevBajkova26} employed the \citet{HR} catalog to analyze the kinematics of open clusters with ages no
greater than 200~Myr using a second-order Bottlinger expansion for angular velocity of Galactic rotation and adopting a solar
Galactocentric distance of $R_0$~=~8.1~$\pm$~0.1~kpc. The above authors also analyzed the variation of the radial cluster
velocity component and of the residual tangential velocity along Galactoicentric radius.

The aim of our study is to analyze the kinematics of the largest sample of young open clusters (with ages $\leq$~100~Myr)
to date --- which also happens to be the most extensive currently available homogeneous sample of young kinematic tracers with
high-precision geometric distances and absolute proper motions --- in order to jointly determine the form of the Galactic rotation
curve and the perturbations (non-circular motions) induced by spiral density waves in terms of the linear theory developed
by \citet{LinShu1, LinShu2}. Unlike our similar study based on Galactic maser data \citep{Rastorguev17},
we do not attempt to refine the underlying  distance scale of open clusters (based on Gaia DR3 trigonometric parallaxes).
For the sample considered, this cannot be achieved with  acceptable precision due to the insufficient reliability and large errors
of the radial velocity estimates for young clusters. The problem is that these estimates are based on radial velocity measurements
of cluster members, which in most the cases are hot B- and early A-type stars. The Gaia RVS spectra of such stars lack the
narrow lines of the calcium triplet, forcing the use of very broad and extremely shallow hydrogen Paschen lines instead.
As a result, the reported errors of the mean radial velocities of young clusters are comparable to the velocity dispersion,
and the unavoidable inaccuracies in estimating these errors lead to very substantial uncertainties in the final evaluation
of the distance scale correction error. For the same reason, we also abandoned the attempt to determine the solar
Galactocentric distance and treat this quantity  as very precisely known, adopting the estimate
of $R_0$~=~8.277~$\pm$~0.009~(statistical  error)~$\pm$~0.030~(systematic error)~kpc, as inferred  from an analysis of the motion
the star S2 orbiting the supermassive black hole at the Galactic center \citep{GRAVITY}.

\section{INITIAL DATA}

As noted above, our data source is the \citet{HR} catalog of open clusters, from which we select objects with ages 100~Myr and
younger. The catalog contains a total of 2396 such clusters, all of them with known mean parallaxes, ages, and proper motions.
Mean radial velocity estimates are available for 1750 of these systems. Figs 1a and 1b show the distributions of proper-motion
component errors for all 2396 young clusters and  Fig. 2, the distribution of radiaal-velocity errors for 1750 clusters.
Figs 3a and 3b  show the distributions of the transversal-velocity component errors for all the clusters considered.
As is evident from the figures, the errors of transversal velocity components are much smaller that the characteristic
velocity dispersion values for young objects of the Milky-Way thin
disk  ($\sigma V_R \sim$~10~km/s, $\sigma V_\theta \sim$~5~km/s - \citep{Rastorguev17}), whereas errors of cluster
mean radial velocities are comparable to the velocity dispersion and even significantly greater for many clusters.
That is why we did not attempt to calibrate the adopted cluster distance scale via the method of statistical parallax.
Deriving an unbiased correction to the distance scale requires bona-fide standard error estimates for radial velocities, whereas in this particular case, the existing error estimates of the mean $V_R$ values are far from reliable.

\begin{figure*}
\includegraphics[scale=0.25]{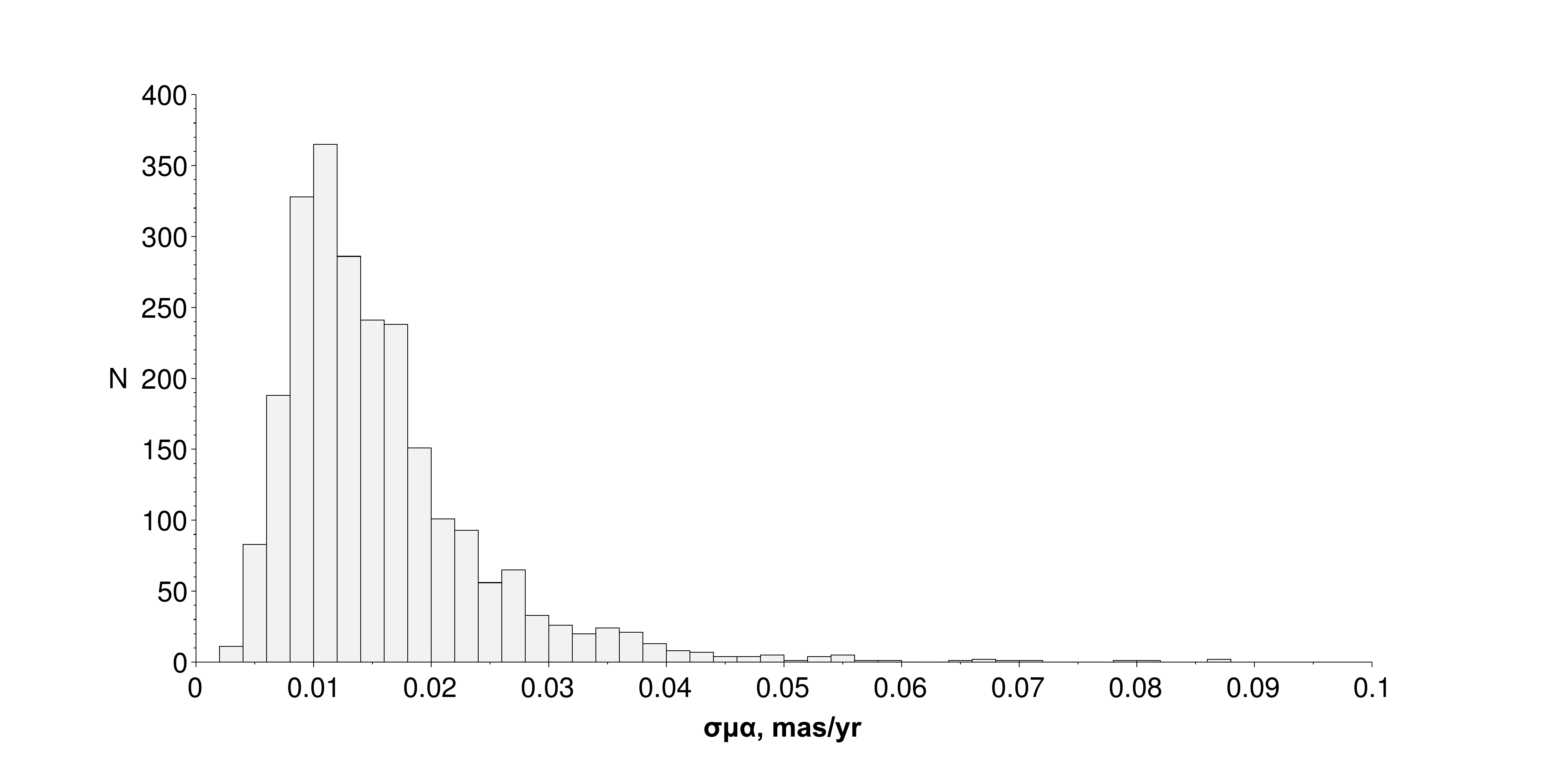}
\includegraphics[scale=0.25]{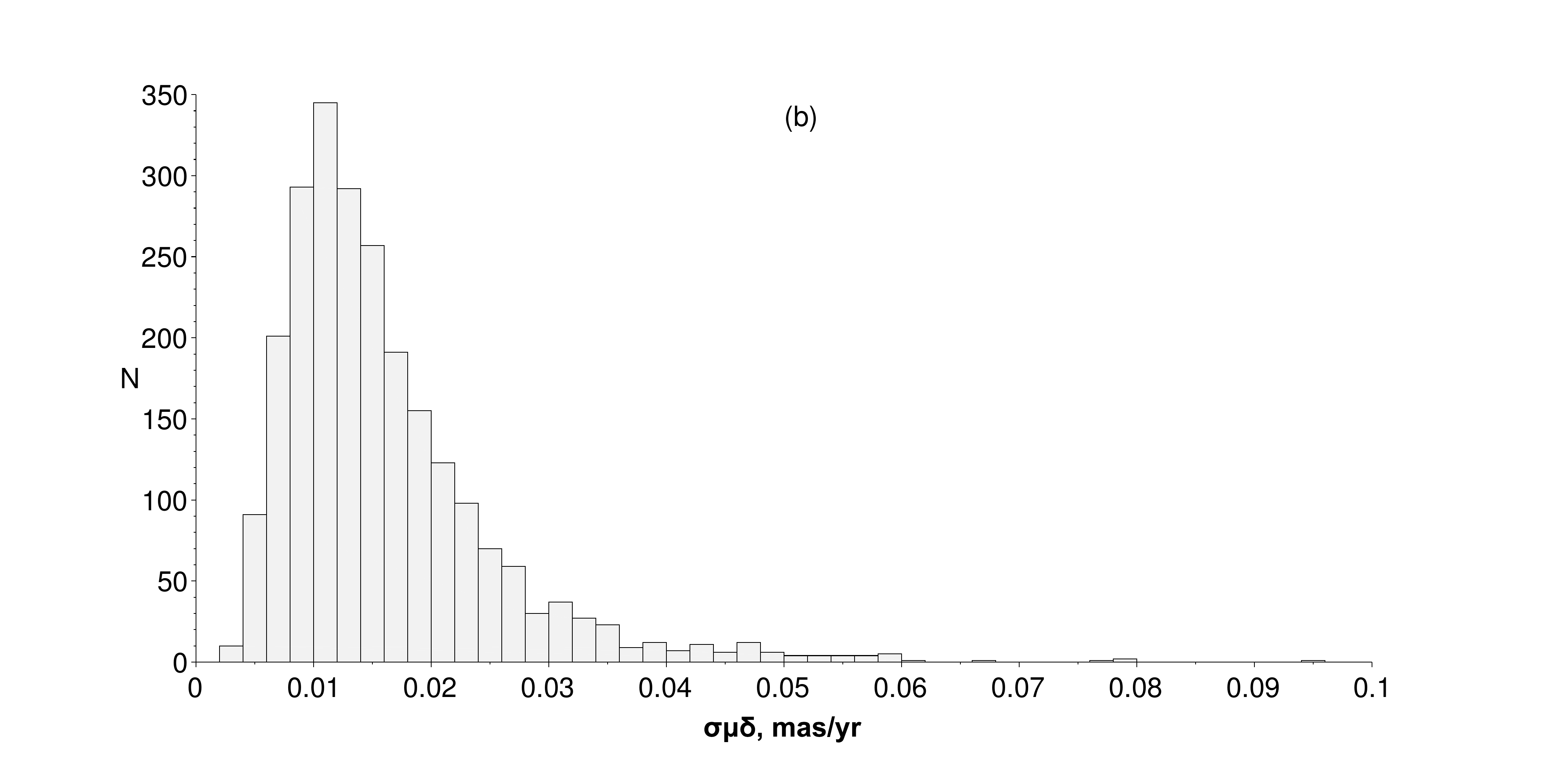}
\caption{Distribution of proper-motion component errors in right ascension (a) and declination (b) for clusters with ages $\leq$~100~Myr.}
\label{PMERR}
\end{figure*}

\begin{figure*}
	\includegraphics[scale=0.25]{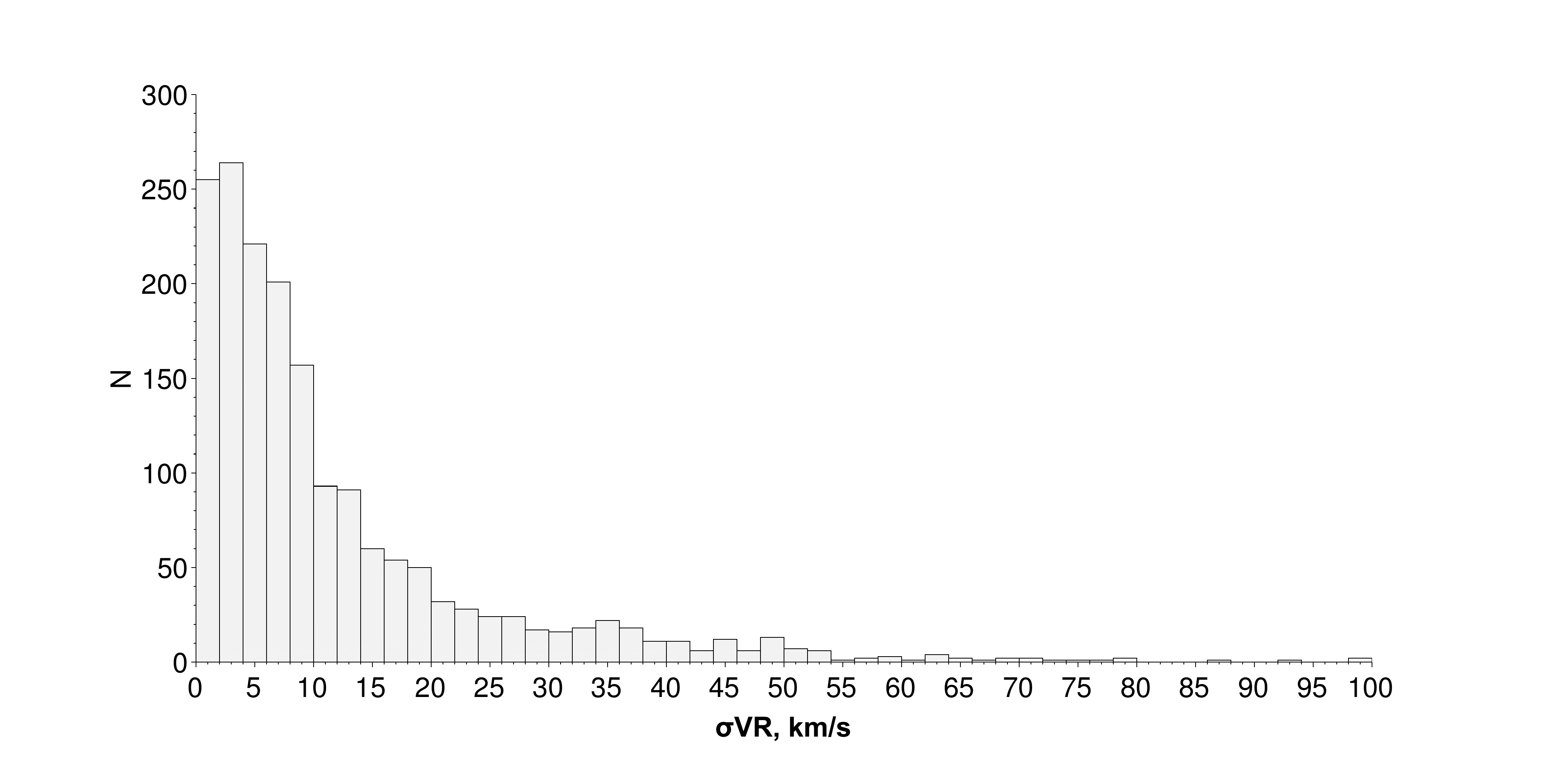}
	\caption{Distribution of errors of mean radial velocities for clusters with ages $\leq$~100~Myr.}
	\label{PMVR}
\end{figure*}

\begin{figure*}
	\includegraphics[scale=0.25]{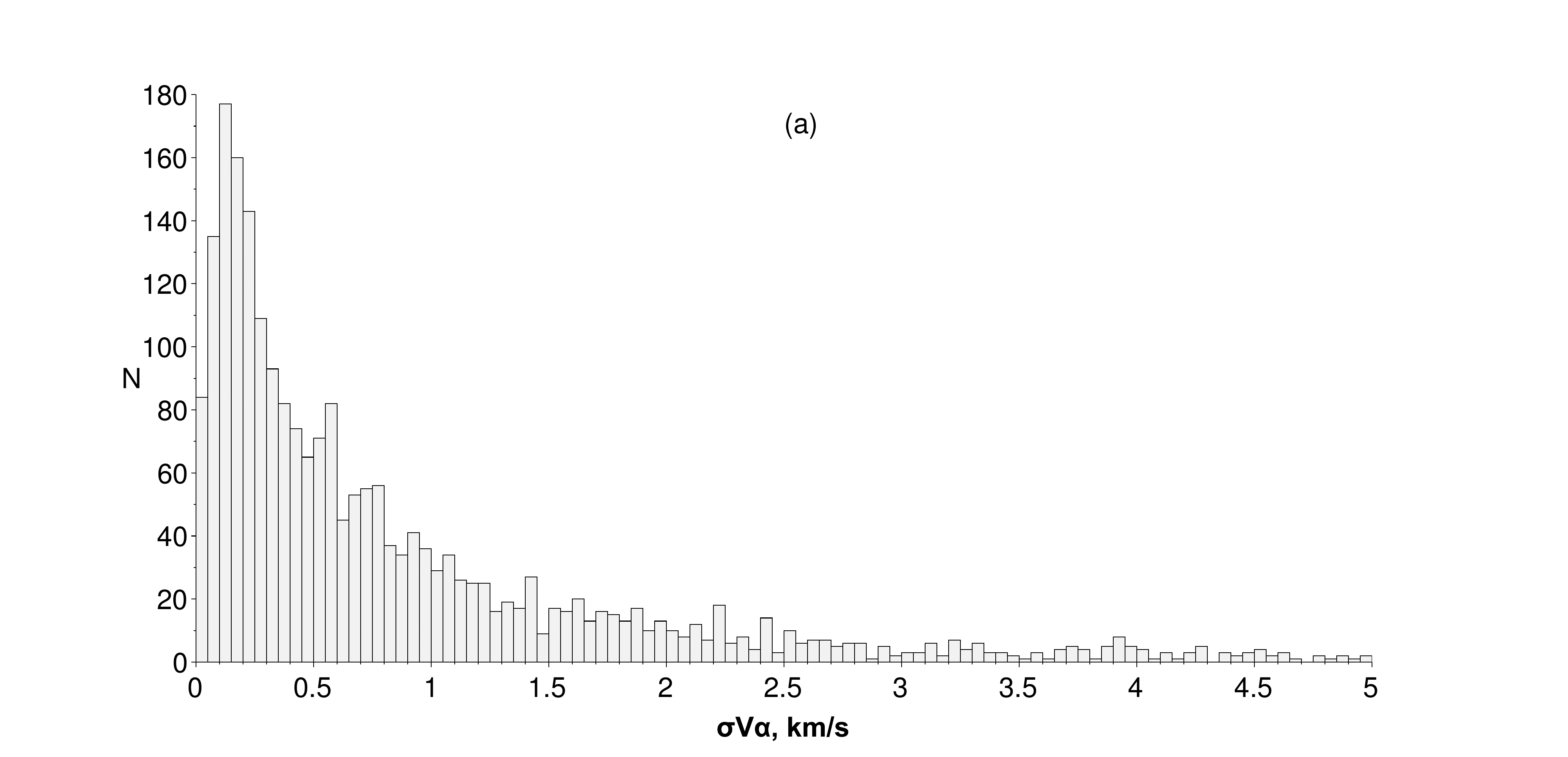}
	\includegraphics[scale=0.25]{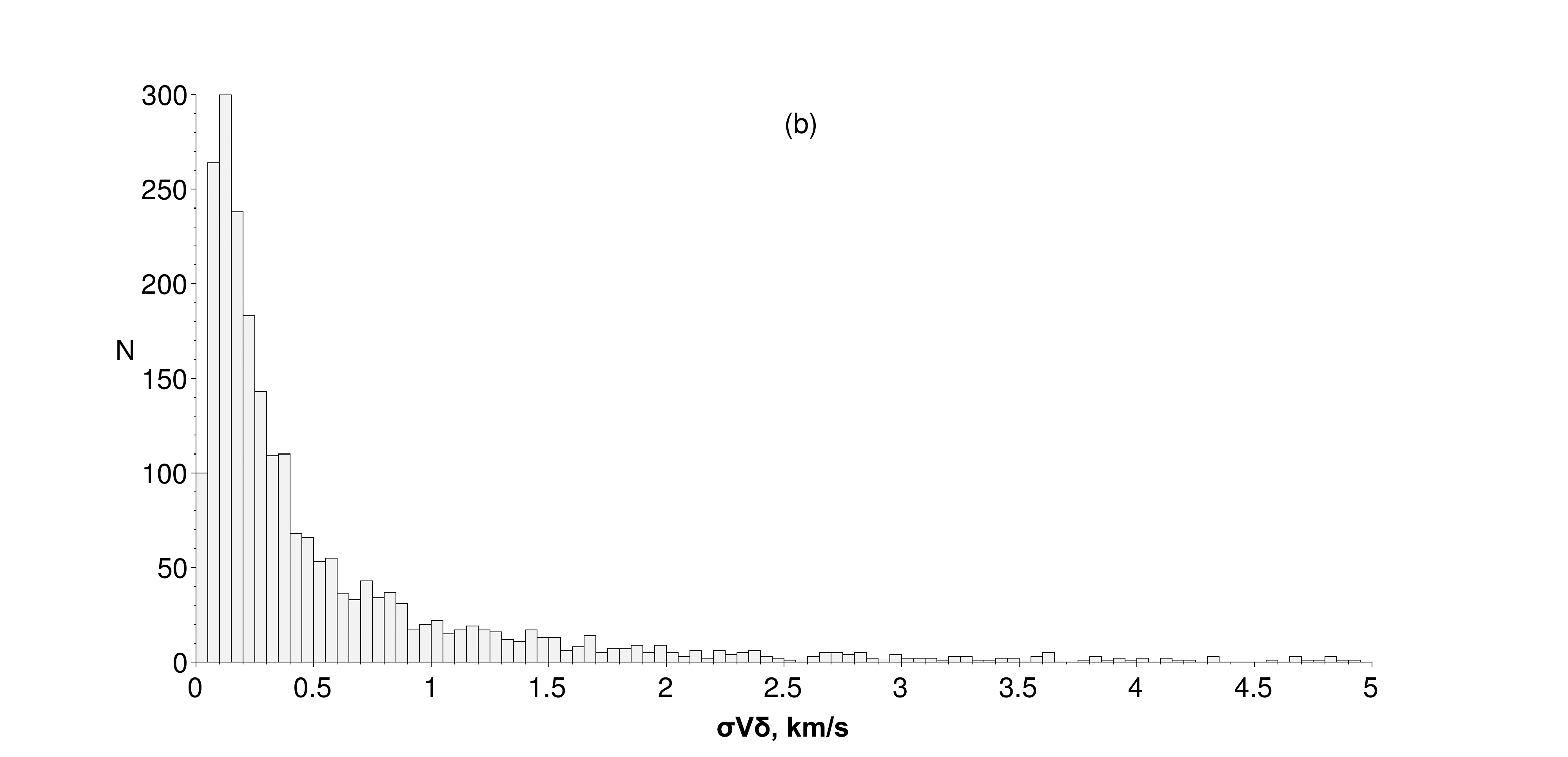}
	\caption{Distribution of errors of right-ascension (a) and declination (b) transversal velocity components ($\sqrt{(\sigma \mu d)^2  + (\mu \sigma d)^2}$) of open clusters with ages $\leq$~100~Myr.}
	\label{PMERR}
\end{figure*}

\section{THE METHOD}

Here we use version C of the statistical parallax method (with constant  vertical and radial components of cluster velocity dispersion
along the Galactocentric radius, and the azimuthal velocity dispersion calculated from the radial velocity dispersion via the
Lindblad relation), as described in detail in our previous paper \citet{Rastorguev17}, without estimating a correction to the
adopted distance scale and without refining the solar Galactocentric distance. The adopted distance scale is based on the
Gaia DR3 trigonometric parallaxes and is considered a priori correct, while the distance from the Sun to the Galactic center
is assumed to be known with very high precision ---
$R_0$~=~8.277~$\pm$~0.009~(statistical  error)~$\pm$~0.030~(systematic error)~kpc \citep{GRAVITY}.
Furthermore, given that most of the clusters in our sample are located within the Galactocentric distance interval from 5 to 13 kpc
(see the histogram in Fig. 4), where the rotation curve of our Galaxy is almost to flat \citep{Rastorguev17},
instead of the Bottlinger expansion for the angular velocity of Galactic rotation  [equation (19) in \citet{Rastorguev17}],
we expand the dependence of the linear Galactic rotation velocity into powers of $R_0/R_g$:
\begin{eqnarray}
	V_{rot}=a_0 + a_1\times (R0/R_g) + a_2\times (R0/R_g)^2 \nonumber\\
	+ a_3\times (R0/R_g)^3\nonumber.
\end{eqnarray}

To avoid high correlation between the parameters we use an equiivalent expansion into a polynomial series in $x=(R_0/R_g)$~--1.0:
\begin{eqnarray}
	V_{rot}=V_{0,rot} + b_1 \times x + ( b_2 \times x^2 ) \nonumber\\
	+ (b_3 \times x^3 )\nonumber,
\end{eqnarray}

where $V_{0,rot}$ is the linear velocity of Galactic rotation at the solar Galacticocentric distance $R_0$, and the
parameters $b_1$, $b_2$ and $b_3$ characterize the derivatives of the rotation curve at this distance. Thus,
the sought parameters of the rotation of the system of young open clusters are the quantities
$V_{0,rot}$, $b_1$, $b_2$, $b_3$, which describe the form of the Galactic rotation curve, the components
$\sigma V_R$ and $\sigma V_Z$ of the dispersion tensor of the residual velocities, the amplitudes $f_R$ and $f_{\Theta}$
of the radial and tangential velocity components caused by the spiral wave, the phase of the Sun in the spiral wave $\chi_0$,
the pitch angle $i$ of the spiral
perturbations,  and the velocity components $U_0$, $V_0$ and $W_0$ of the local sample relative
to the Sun - a total of 13 parameters.
The $x$ quantity is dimensionless and therefore the parameters $b_1$, $b_2$, $b_3$, as well as $V_{0,rot}$ have the dimension km/s.

\begin{figure*}
	\includegraphics[scale=0.25]{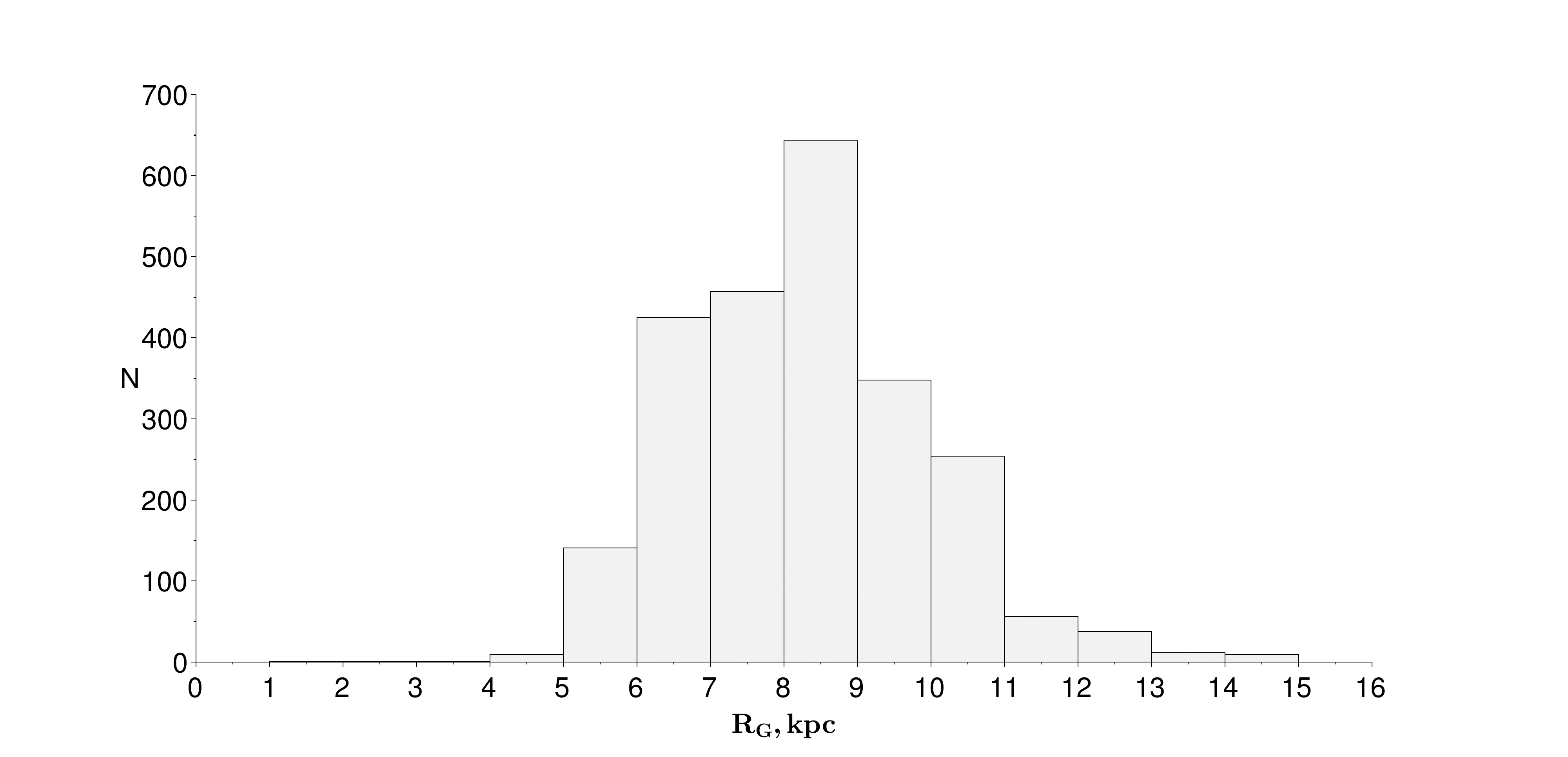}
	\caption{Distribution of Galactoaxial distances of open clusters with ages $\leq$~100~Myr.}
	\label{RG}
\end{figure*}

Like in our maser kinematics study \citep{Rastorguev17}, here we use the maximum likelihood method to determine the values
of the target parameters. According to \citet{Murray}, the likelihood function for the three-dimensional distribution of
differences between the observed and model velocities of objects (in this case, clusters) can be written as
\[
f(\delta \vec V_{loc} \mid \Lambda) =
\]
\[ = (2\pi )^{-3/2}  \cdot \left|
{L_{loc}} \right|^{ - 1/2}  \cdot \exp \{ -{1 \over 2} \cdot
\delta \vec V_{loc}^T \times L_{loc}^{-1}\times \delta
\vec V_{loc}\} ,
\] 

Here, $\Lambda$ is a set of unknown parameters describing the model velocity field (including the velocity dispersion tensor).
In the case of a two-dimensional observed velocity field (where the proper motions are considered known, while the radial
velocity information is absent or not utilized), the formula for the likelihood function has the same form.

As mentioned in the Introduction, the mean cluster radial velocities and the estimates of their standard errors for young clusters
are unreliable, and therefore we infer the sought parameters by minimizing the two-dimensional likelihood function via the NEWUOA
multidimensional minimization method developed by Powell \cite{Powell} as implemented by \citet{Siromakha15}
and Markov Chain Monte Carlo (MCMC) method. We determine the matrix of parameter covariances via the
Markov Chain Monte Carlo (MCMC) method using  the affine-invariant sampler proposed by \citet{GW}. To mitigate the effect
of eventual gross errors and outliers, we apply an analogue of the so-called three-sigma rule --- namely, we exclude clusters
for which twice the exponent value in the likelihood function for the found solution exceeds 11.83 (for a two-dimensional
normal distribution, the probability of such deviations is 1 - 0.997 = 0.003, which corresponds to a deviation of more than
three standard deviations in the one-dimensional case). We apply this process iteratively: we compute the solution for the
entire sample, exclude outlier clusters, repeat the entire procedure for the remaining sample, and so on, until no outliers remain.

To further validate the results obtained, we repeated our computations with the 3D version of residual velocity distribution for the
sample of clusters with available estimates of mean radial velocities. We do it using two other formulas for describing the form of
the rotation curve, $V_{rot}$~=~$\Omega(R_g) \times R_g$:

\begin{equation}
	\Omega= (A/R_g) + (B/R_g^2 ) + C,
\end{equation}

and 

\begin{eqnarray}
	\Omega= (A/R_g) + (B/R_g^2 ) + (C/R_g^3 ) \nonumber\\
	+ D + (E \times R_g ) + (F \times R_g^2).
\end{eqnarray}	

To see how stable is the adopted method we also repeated the computations based on proper motions exclusively
without using radial velocities.

\section{RESULTS AND DISCUSSION}

Table~\ref{res_ml} presents the parameter values and their standard errors inferred by the maximum likelihood method,
while Table~\ref{res_mcmc} shows the parameter values and errors determined by the Markov Chain Monte Carlo method.
Fig. 5 shows the one-dimensional histograms and corner plots that visually illustrate the marginal and joint pairwise
distributions of each parameter.

\subsection{Rotation Curve}

Fig. 6 shows the resulting rotation curve for the sample of clusters from \citet{HR} with ages up to 100 Myr inclusive
(the red curve) compared with the rotation curve obtained by \citet{Rastorguev17} from Galactic maser data (the blue dashed line).
Overall, the curves agree very well with each other; the main difference is a flatter shape of the maser-based curve compared
to that inferred from young clusters. The latter exhibits a slow, steady decline beyond Galactoaxial distances of
\(R_g \sim 7.5\text{--}8\) kpc and a steeper drop at Galactoaxial distances of less than 6 kpc. We also show the rotation curve
implied by the parameters inferred by \citet{BobylevBajkova26} (the dashed magenta curve), which runs systematically below
both our maser- and open-cluster-based rotation curves. However, this appears to be mostly due to the difference in the
adopted solar Galactocentric distance. When recomputed and rescaled with $R_0$~=~8.277~kpc, the \citet{BobylevBajkova26}
rotation curve agrees quite well with our results.

We find the linear Galactic rotation velocity at the solar Galactocentric distance to be $V_{0,rot}$~=~242~$\pm$~1~km/s, and,
as is evident from Tables~\ref{res_ml}--\ref{res_mcmc2}, practically does not depend on the adopted formula for the rotation
curve and on whether we use the 2D or 3D velocity dispersion model. These estimates are slightly greater than but still
consistent with our maser-based estimates (235--238~$\pm$~7~km/s) and with the recent  estimate by \citet{Bobylev_etal25a}
based on the data for 212 masers (240--242~$\pm$~4~km/s), but not with the estimate obtained for young clusters
by \citet{BobylevBajkova26} (235~$\pm$~3~km/s). This again can be partly explained by different adopted values of the solar
Galacticentric distance.

\subsection{Parameters of the Spiral Perturbation}

As is evident from the results presented in Tables~\ref{res_ml}--\ref{res_mcmc2}, we obtained bona fide estimates for the
parameters of velocity-field perturbation caused by spiral density wave. Like in the case of the linear (and angular)
velocity of Galactic rotation at the solar distance, the choice of the formula adopted for the rotation curve and the
use of either the 2D or 3D velocity distribution model have virtually no effect on the results.
As is evident from Tables~1 and 2, our analysis yields bona fide estimates for the parameters describing velocity field
perturbations induced by the spiral density wave. We find significant amplitudes for both the radial and tangential
velocity perturbations. Our tangential perturbation amplitude ($f_{\theta}$~=~-3.1~$\pm$~0.4~km/s) is quite consistent
with the result of our maser-based analysis ($f_{\theta}$~=~-2.8~$\pm$~1.0~km/s - the omission of the minus sign in
\citet{Rastorguev17} is a typo), although the radial perturbation amplitude ($f_R$~=~-3.0~$\pm$~0.4~km/s)
significantly differs from our estimate inferred from maser kinematics ($f_R$~=~-7.0~$\pm$~1.5~km/s). Both amplitudes
in our open-cluster solution differ from recent maser-based estimates study
by \citet{Bobylev_etal25a}  ($f_R$, $f_{\theta}$)~=~(-6.8~$\pm$~1.0, +0.8~$\pm$~1.0)~km/s.

Our estimate of the pitch angle of the spiral pattern presented in Tables~\ref{res_ml} and \ref{res_mcmc} and inferred from
a large sample of young clusters ($i$~=~-11.69~$\pm$0.33$\degr$) differs from the results obtained for the sample of clusters
with available estimates of mean radial velocities and presented in Tables~\ref{res_mcmc1}--\ref{res_mcmc2}
($i$~=~-10.4--~-10.7$\degr$), which, in turn, are consistent with the value inferred from masers ($i$~=~-10.39~$\pm$0.25$\degr$).
The latter differs from the value inferred from the sample of all young clusters with known proper motions , possibly,
because the two tracer samples cover different areas and because spiral arms are not ideal logarithmic spirals. However,
it is consistent with the estimates obtained by \citet{BobylevBajkova22} from an analysis of the distribution of classical
Cepheids in three Galactic spiral arms ($i$~=~-12.0~--~-12.7~$\degr$). Our pitch-angle estimate is also  consistent with the
results obtained by \citet{Tang} based on an analysis of the kinematics of red-giant-branch stars ($i$~=~-8.0~--~-12.6~$\degr$).
Our estimate for the phase of the Sun in the spiral arm ($\chi_0$~=~138~$\pm$~5~$\degr$) is consistent with our earlier
maser-based result ($\chi_0$~=~123--130~$\pm$~11~$\degr$) and with the recent $\chi_0$~=~140~$\degr$
estimate by \citet{Bobylev_etal25a}.

\subsection{Local Solar Velocity and Solar Motion with Respect to the Galactic Center of Mass}

Our inferred components of the local solar velocity
($U_0$, $V_0$, $W_0$)~=~(-9.51~$\pm$~0.41, -11.19~$\pm$~0.56, -7.90~$\pm$~0.09)~km/s derived without using radial-velocity data
agree quite well with the values determined by \citet{Robin} in terms of a self-consistent dynamical model of
the Milky Way disk adjusted to Gaia data [($U_0$, $V_0$, $W_0$)~=~(-10.59~$\pm$~0.56, -11.06~$\pm$~0.94, -7.66~$\pm$~0.43)~km/s]
and more or less agree with the open-cluster-based solution
of \citet{Bobylev_etal25a} [($U_0$, $V_0$, $W_0$)~=~(-8.9~--~-9.1~$\pm$~0.2, -8.6~--~-12.2~$\pm$~0.3, -7.8~--~-8.0~$\pm$~0.2)~km/s].
Our $U_0$ and $W_0$ components are also quite consistent with the results of our maser-based solution
[($U_0$, $W_0$)~=~(-10.98~$\pm$~1.40,  -8.93~$\pm$~1.05)~km/s \citep{Rastorguev17}], however, the $V_0$ component differs
significantly ($V_0$~=~-19.62~$\pm$~1.15~km/s \citep{Rastorguev17}), but is in line with those obtained in a recent study of
maser kinematics by \citet{Bobylev_etal25a} ($V_0$~=~-8.7~--~-12.2~km/s). Note that 3D solution in our case
yields $V_0$~=-12.9--~-13.7~km/s, which also appreciably differs from the result obtained using only proper motions,
although not to the same extent as in our maser-based solution. This discrepancy must be explained by poor quality of
estimates of mean cluster radial velocities and their errors.

Note that $U_0$ and $W_0$ components can be considered to coincide with the corresponding components of the solar velocity relative
to the Galactic center of mass. As for the component of the Galactocetric solar motion in the direction of Galactic rotation,
it can be estimated as $V_0$-$V_{0,rot}$~=253.50~$\pm$~1.11~km/s. This component can also be accurately determined by combining
the very accurate estimate of the solar Galactiocentric distance by \citet{GRAVITY} mentioned above
($R_0$~=~8.277~$\pm$~0.009~$\pm$~0.030~kpc) with very highly accurate measurement of the Sgr~A$^*$ proper motion
by \citet{Reid20} ($\mu_l$(Sgr A$^*$)~=~-6.411~$\pm$~0.008~mas/yr, $\mu_b$(Sgr A$^*$)~=~-0.219~$\pm$~0.007~mas/yr),
implying $V_{0,GC}$~=~251.55~$\pm$~0.98~km/s and $W_{0,GC}$~=~8.59~$\pm$~0.27~km/s. These values differ from our results
by $\sim$~1.3 and 2.4~$\sigma$, respectively, although the absolute differences amount just to 1.95 and 0.69~km/s,
respectively, and so the agreement can be considered marginally satisfactory.

\subsection{Radial and Vertical Components of the Velocity Dispersion Tensor}
	
Our results for the radial and vertical components of the velocity dispersion tensor for all solutions are consistent with the young
age of the clusters considered. Note that our estimates obtained in terms of 2D solutions
($\sigma V_R$, $\sigma V_Z$)~=~(10.6--10.9, 4.6--4.7)~km/s are consistent with those that we inferred earlier from maser data.
The significantly greater radial velocity dispersion inferred in 3D solutions ($\sigma V_R$~=~12.5~km/s) can be explained by
the poor reliability of available estimates of errors of mean cluster radial velocities - they are evidently underestimated
and the extra scatter resulting from this underestimation  ``propagates'' into velocity dispersion. At the same time,
radial-velocity data have virtually no effect on the estimated vertical velocity components and their dispersions because
most of the clusters have small Galactic latitudes, resulting in excellent agreement between the estimates of these parameters
across all solutions.

\begin{figure*}
	\includegraphics[scale=0.25]{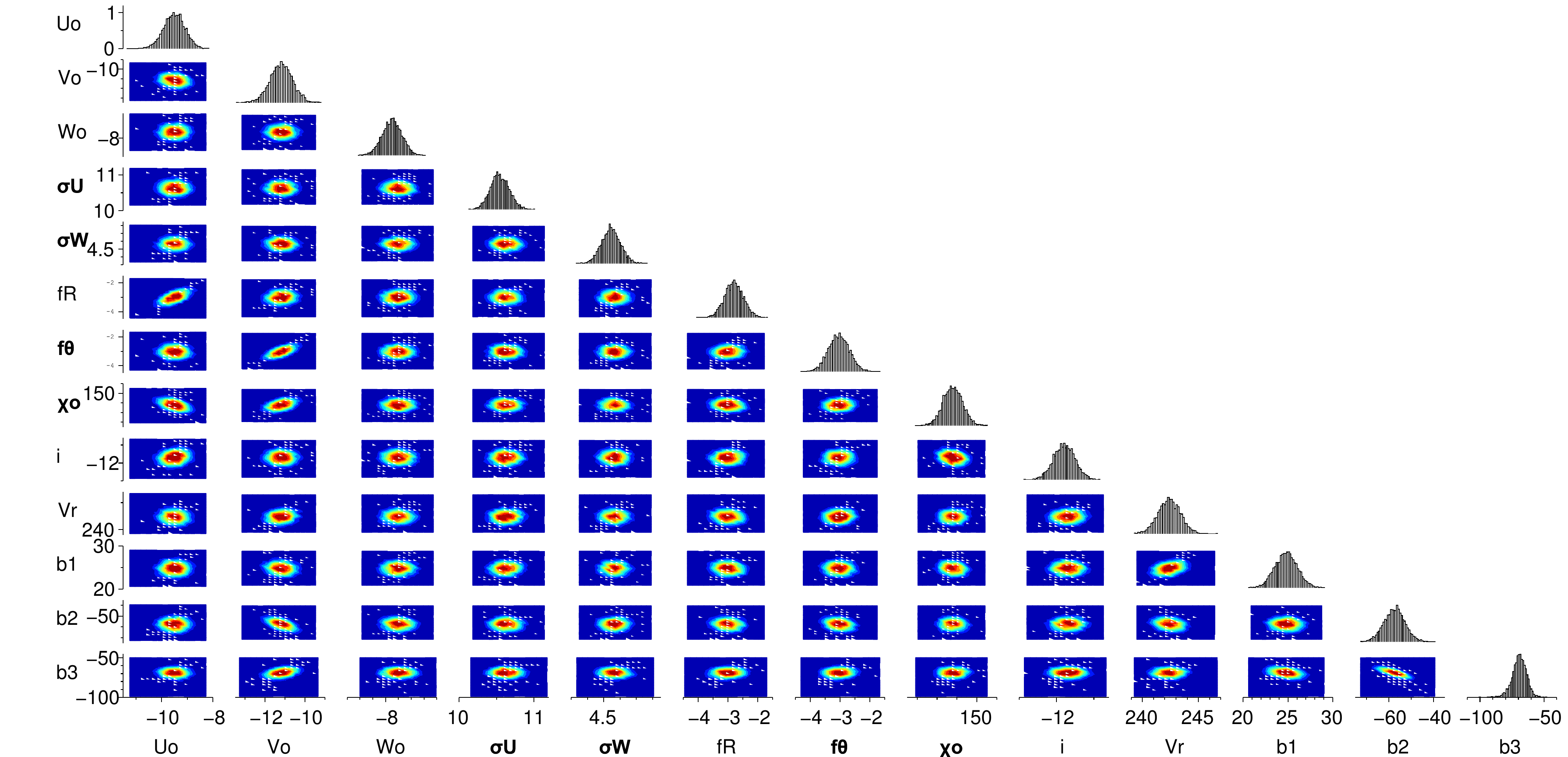}
	\caption{The corner plot  for the parameters of the kinematic model inferred from the data for the sample of open clusters with ages $\leq$~100~Myr.}
	\label{RG}
\end{figure*}

\begin{figure*}
	\includegraphics[scale=0.25]{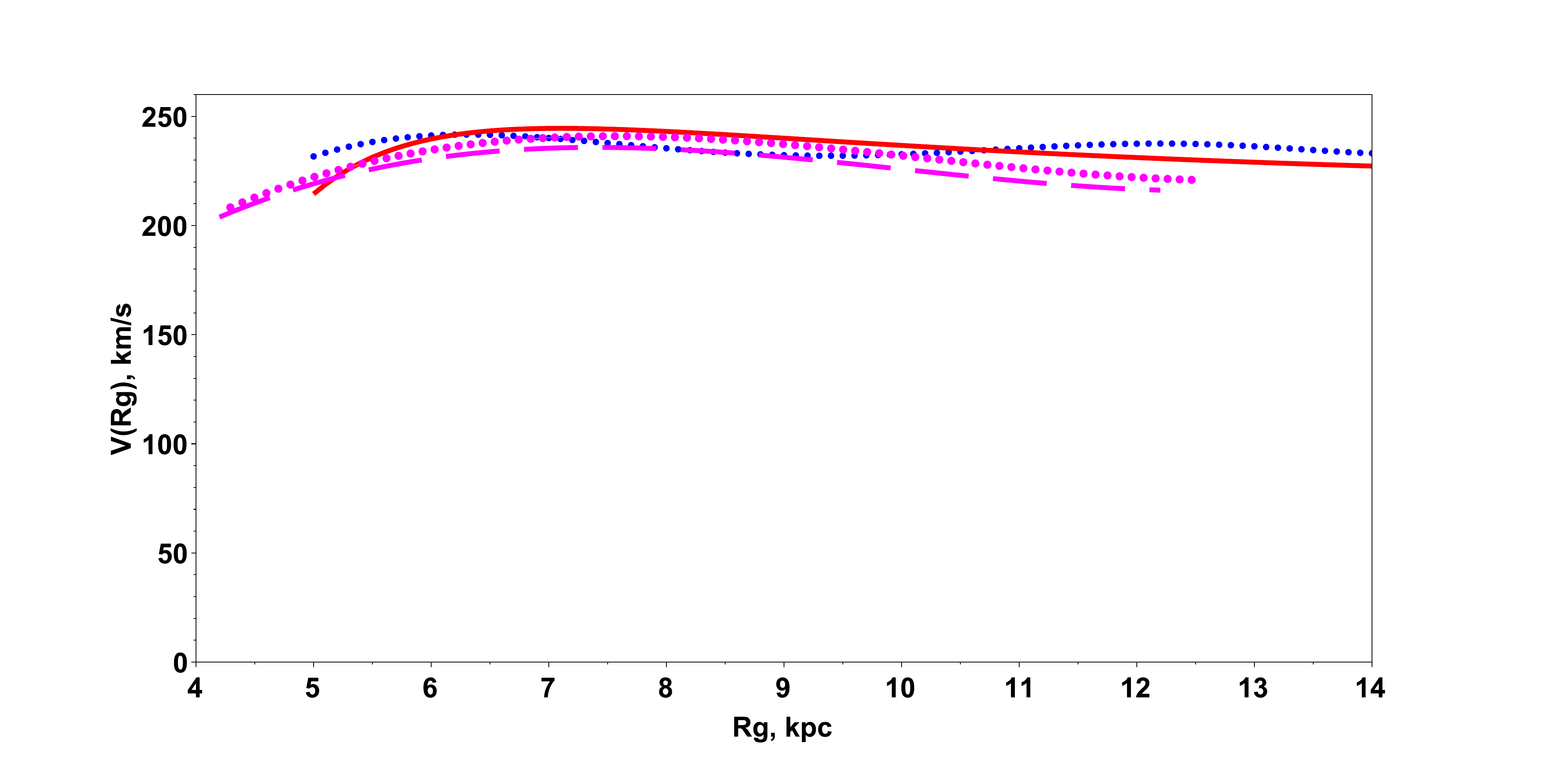}
	\caption{Rotation curve inferred from the data for the sample of open clusters with ages $\leq$~100~Myr (the red curve) according to Table~2 compared with the rotation curve inferred from Galactic maser data (the blue dotted line - model C1 from \citet{Rastorguev17}).}
	\label{RG}
\end{figure*}

\begin{table}
	\caption {Parameter values and their standard errors inferred using the maximum-likelihood method in terms of \textbf{model C1} from the data for 2384 out of 2396 open clusters: circular differential rotation of the disk plus perturbations induced by a four-armed spiral pattern  with constant radial and vertical velocity dispersions $\sigma U$ and $\sigma W$.} \label{res_ml}
	\medskip
	\begin{tabular}{l | r}
		\hline
		$U_0$ , km/s        &  -9.53 $\pm$ 0.41 \\
		$V_0$ , km/s        & -11.21 $\pm$ 0.56 \\
		$W_0$ , km/s        &  -7.90 $\pm$ 0.09 \\
		$\sigma V_R$, km/s  &  10.60 $\pm$ 0.16 \\ 
		$\sigma V_Z$, km/s  &   4.56 $\pm$ 0.07 \\ 
		$V_{0,rot}$ , km/s  & 242.34 $\pm$ 1.04 \\
		$b_1$, km/s         & +24.71 $\pm$ 1.09 \\ 
		$b_2$, km/s         & -57.81 $\pm$ 2.10 \\ 
		$b_3$, km/s         & -68.22 $\pm$ 2.41 \\ 
		$f_R$, km/s         &  -3.03 $\pm$ 0.39 \\ 
		$f_{\theta}$, km/s   &  -3.06 $\pm$ 0.36 \\
		$\chi_0$, deg        & 137.94 $\pm$ 5.09 \\
		$i$, deg             & -11.69 $\pm$ 0.32 \\
		\hline \hline
	\end{tabular}
\end{table}

\begin{table}
	\caption {The parameter values and their standard errors inferred using the Markov chain Monte-Carlo method in terms of  \textbf{model C1} from the data for 2384 out of  2396 open clusters.} \label{res_mcmc}
	\medskip
	\begin{tabular}{l | r}
		\hline
		$U_0$ , km/s        &  -9.51 $\pm$ 0.41 \\
		$V_0$ , km/s        & -11.19 $\pm$ 0.56 \\
		$W_0$ , km/s        &  -7.90 $\pm$ 0.09 \\
		$\sigma V_R$, km/s  &  10.63 $\pm$ 0.15 \\ 
		$\sigma V_Z$, km/s  &   4.56 $\pm$ 0.07 \\ 
		$V_{0,rot}$ , km/s  & 242.35 $\pm$ 1.05 \\
		$b_1$, km/s         & +24.78 $\pm$ 1.32 \\ 
		$b_2$, km/s         & -57.63 $\pm$ 4.93 \\ 
		$b_3$, km/s         & -68.32 $\pm$ 2.41 \\ 
		$f_R$, km/s         &  -3.03 $\pm$ 0.39 \\ 
		$f_{\theta}$, km/s   &  -3.06 $\pm$ 0.36 \\
		$\chi_0$, deg        & 137.93 $\pm$ 5.16 \\
		$i$, deg             & -11.69 $\pm$ 0.33 \\
		$\Omega$,     km s$^{-1}$ kpc$^{-1}$ &  29.280 $\pm$ 0.127 \\			
		$d\Omega/dR$, km s$^{-1}$ kpc$^{-2}$ &  -3.900 $\pm$ 0.028 \\						
		$A$ , km s$^{-1}$ kpc$^{-1}$         &  16.137 $\pm$ 0.118 \\			
		$B$ , km s$^{-1}$ kpc$^{-1}$         & -13.143 $\pm$ 0.083 \\			
		\hline \hline
	\end{tabular}
\end{table}

\begin{table}
	\caption {The parameter values and their standard errors inferred using the Markov chain Monte-Carlo method in terms of 2D and 3D versions of \textbf{model C1} and angular-velocity expansion formula~(1) from the data for 1704 open clusters with available mean radial-velocity estimates.		
		} \label{res_mcmc1}
	\medskip
	\begin{tabular}{l | r |  r}
		\hline
		Parameter             &      2D            & 3D \\
		$U_0$ , km/s        &  -9.82 $\pm$ 0.81  & -9.65 $\pm$ 0.90\\
		$V_0$ , km/s        & -10.15 $\pm$ 0.76  & -12.93 $\pm$ 0.86\\
		$W_0$ , km/s        &  -7.89 $\pm$ 0.31  & -7.95 $\pm$ 0.34\\
		$\sigma V_R$, km/s  &  10.91 $\pm$ 0.88  & 12.54 $\pm$ 0.92\\ 
		$\sigma V_Z$, km/s  &   4.67 $\pm$ 0.36  & 4.64 $\pm$ 0.34\\ 
		$V_{0,rot}$ , km/s  &   240.0 $\pm$ 1.7     & 239.8 $\pm$ 1.7\\
		$\Omega_0$, km/s/kpc&   29.00 $\pm$ 0.20       & 28.97 $\pm$ 0.20\\ 
		$f_R$, km/s         &  -2.73 $\pm$ 0.74  & -2.91 $\pm$ 0.83\\ 
		$f_{\Theta}$, km/s  &  -2.87 $\pm$ 0.80  & -2.95 $\pm$ 0.88\\
		$\chi_0$, deg        & 140.47 $\pm$ 12.68 & 136.83 $\pm$ 14.16\\
		$i$, deg             & -10.71 $\pm$ 0.79  & -10.43 $\pm$ 0.81\\
		\hline \hline
	\end{tabular}
\end{table}

\begin{table}
	\caption {The parameter values and their standard errors inferred using the Markov chain Monte-Carlo method in terms of 2D and 3D versions of \textbf{model C1} and angular-velocity expansion formula~(2) from the data for 1704 open clusters with available mean radial-velocity estimates.} \label{res_mcmc2}
	\medskip
	\begin{tabular}{l | r |  r}
		\hline
		Parameter             &      2D            & 3D \\
		$U_0$ , km/s        &  -9.83 $\pm$ 0.49  & -9.60 $\pm$ 0.49\\
		$V_0$ , km/s        & -10.35 $\pm$ 0.45  & -13.65 $\pm$ 0.45\\
		$W_0$ , km/s        &  -7.91 $\pm$ 0.20  & -7.95 $\pm$ 0.19\\
		$\sigma V_R$, km/s  &  10.53 $\pm$ 0.46  & 12.52 $\pm$ 0.48\\ 
		$\sigma V_Z$, km/s  &   4.65 $\pm$ 0.21  & 4.64 $\pm$ 0.19\\ 
		$V_{0,rot}$ , km/s  &  241.0 $\pm$  1.7     & 238.3 $\pm$ 1.7\\
		$\Omega_0$, km/s/kpc&  29.12 $\pm$  0.20     & 28.79 $\pm$ 0.20\\ 
		$f_R$, km/s         &  -2.87 $\pm$ 0.50  & -2.83 $\pm$ 0.49\\ 
		$f_{\Theta}$, km/s  &  -2.65 $\pm$ 0.48  & -3.14 $\pm$ 0.49\\
		$\chi_0$, deg        & 142.03 $\pm$ 8.28 & 136.64 $\pm$ 7.22\\
		$i$, deg             & -10.83 $\pm$ 0.49  & -10.80 $\pm$ 0.49\\
		\hline \hline
	\end{tabular}
\end{table}

\section{CONCLUSIONS}
We applied the maximum-likelihood method to a sample of 2396 young open clusters (with ages $\leq$~100~Myr) spanning
the Galactocentric distance interval from 1.7 to 15.2~kpc to study the parameters of their velocity field
assuming that velocity dispersion is constant in the radial and vertical directions.
We find the mean heliocentroic velocity components of sample clusters to be
($U_0$, $V_0$, $W_0$)~=~(-9.51~$\pm$~0.41, -11.19~$\pm$~0.56, -7.90~$\pm$~0.09)~km/s and traced the rotation curve over
the Galactocentric distance interval from 5 to 14~kpc. Our rotation curve proved to be overall consistent with most recent results
obtained by other authors, with linear velocity at the solar Galactocetric radius equal to  $V_{0,rot}$~=~242.35~$\pm$~1.05~km/s,
in line with most of the recent results, and a slow decline with
increasing Galactocentric distance starting from $\sim$~7.5--8~kpc. We find the radial and vertical velocity dispersions
of the cluster sample considered to be ($\sigma U$, $\sigma W$)~=~(10.63~$\pm$~0.15,4.56~$\pm$~0.07)~km/s.

The pitch angle of the kinematical spiral pattern outlined by young open clusters appears to be $i$~=~-11.69~$\pm$~0.33$\degr$
and the phase of the Sun in the spiral wave, $\chi_0$~=~137.93~$\pm$~5.2$\degr$, and are also consistent with recent estimates.
We infer the amplitudes of the spiral-pattern-induced radial and tangential perturbations
to be ($f_R$, $f_{\theta}$)~=~(-3.03~$\pm$~0.39, -3.06~$\pm$~0.36)~km/s.

\begin{acknowledgments}
The authors acknowledge the use of the VizieR catalogue access tool, 
provided by the Centre de Données astronomiques de Strasbourg (CDS), France.
\end{acknowledgments}

\section*{FUNDING}
The study was conducted under the state assignment of Lomonosov Moscow State University.

\section*{CONFLICT OF INTEREST}
The authors declare no conflicts of interest.

\bibliographystyle{aspb1}
\bibliography{Rastorguev_en}

\end{document}